\documentclass[twocolumn,showpacs,preprintnumbers]{revtex4}
\usepackage{amsmath}
\usepackage{amssymb}
\usepackage{amsfonts}
\usepackage{graphicx}
\usepackage{dcolumn}
\usepackage{bm}

\input{tcilatex}

\begin{document}

\title{Maximum Entropy, \\
the Collective Welfare Principle\\
and the Globalization Process}
\author{Esteban Guevara Hidalgo$^{\dag \ddag }$}
\affiliation{$^{\dag }$Departamento de F\'{\i}sica, Escuela Polit\'{e}cnica Nacional,
Quito, Ecuador\\
$^{\ddag }$SI\'{O}N, Autopista General Rumi\~{n}ahui, Urbanizaci\'{o}n Ed%
\'{e}n del Valle, Sector 5, Calle 1 y Calle A \# 79, Quito, Ecuador}

\begin{abstract}
Although both systems analyzed are described through two theories apparently
different (quantum mechanics and game theory) it is shown that both are
analogous and thus exactly equivalents. The quantum analogue of the
replicator dynamics is the von Neumann equation. A population can be
represented by a quantum system in which each subpopulation playing strategy 
$s_{i}$ is represented by a pure ensemble in state $\left\vert \Psi
_{k}(t)\right\rangle $ and with probability $p_{k}$. The probability $x_{i}$
of playing strategy $s_{i}$ or the relative frequency of the individuals
using strategy $s_{i}$ in that population can be represented as the
probability $\rho _{ii}$ of finding each pure ensemble in the state $%
\left\vert i\right\rangle $.

Quantum mechanics could be used to explain more correctly biological and
economical processes. It could even encloses theories like games and
evolutionary dynamics. We can take some concepts and definitions from
quantum mechanics and physics for the best understanding of the behavior of
economics and biology. Also, we could maybe understand nature like a game in
where its players compete for a common welfare and the equilibrium of the
system that they are members. All the members of our system will play a game
in which its maximum payoff is the equilibrium of the system. They act as a
whole besides individuals like they obey a rule in where they prefer to work
for the welfare of the collective besides the individual welfare. A system
where its members are in Nash Equilibrium (or ESS) is exactly equivalent to
a system in a maximum entropy state. A system is stable only if it maximizes
the welfare of the collective above the welfare of the individual. If it is
maximized the welfare of the individual above the welfare of the collective
the system gets unstable an eventually collapses. The results of this work
shows that the \textquotedblleft globalization\textquotedblright\ process
has a behavior exactly equivalent to a system that is tending to a maximum
entropy state and predicts the apparition of big common markets and strong
common currencies that will find its \textquotedblleft
equilibrium\textquotedblright\ by decreasing its number until they get a
state characterized by only one common currency and only one common market
around the world.
\end{abstract}

\maketitle
\email{esteban\_guevarah@yahoo.es}

\section{Introduction}

With the purpose of beginning with our analysis we will start from the fact
that a physical system is modeled through quantum mechanics and a
socioeconomical system through evolutionary game theory. The relationships
between these two systems are analyzed and it is shown how although both
systems are described through two theories apparently different both are
analogous and thus exactly equivalents.

\bigskip

The quantum analogue of the replicator dynamics is the von Neumann equation.
The natural trend of these systems is to a maximum entropy state which is
their state of equilibrium. A system in where all its members are in Nash
equilibrium is equivalent to a system in a maximum entropy state. The
\textquotedblleft globalization\textquotedblright\ process has a behavior
exactly equivalent to a system that is tending to a maximum entropy state
and it is predicted the apparition of big common markets and strong common
currencies that will reach the \textquotedblleft
equilibrium\textquotedblright\ by decreasing its number until they get a
state characterized by only one common currency and one common market around
the world.

\bigskip

We could understand nature like a game in where its players compete for a
common welfare and the equilibrium of the system that they are members. All
the members of our system will play a game in which its maximum payoff is
the equilibrium of the system. They act as a whole besides individuals like
they obey a rule in where they prefer to work for the welfare of the
collective besides the individual welfare. The stability of the system is
given by the maximization of the welfare of the collective. If it is
maximized the welfare of the individual above the welfare of the collective
the system gets unstable an eventually collapses.

\section{The Replicator Dynamics \& Evolutionary Game Theory}

Game theory \cite{1,2,3} is the study of decision making of competing agents
in some conflict situation. It tries to understand the birth and the
development of conflicting or cooperative behaviors among a group of
individuals who behave rationally and strategically according to their
personal interests. Each member in the group strive to maximize its welfare
by choosing the best courses of strategies from a cooperative or individual
point of view.

\bigskip

The central equilibrium concept in game theory is the Nash Equilibrium. A 
\textit{Nash equilibrium }(NE) is a set of strategies, one for each player,
such that no player has an incentive to unilaterally change his action.
Players are in equilibrium if a change in strategies by any one of them
would lead that player to earn less than if he remained with his current
strategy. A \textit{Nash equilibrium} satisfies the following condition%
\begin{equation}
E(p,p)\geq E(r,p)\text{,}  \label{1}
\end{equation}%
where $E(s_{i},s_{j})$ is a real number that represents the payoff obtained
by a player who plays the strategy $s_{i}$\ against an opponent who plays
the strategy $s_{j}$. A player can not increase his payoff if he decides to
play the strategy $r$ instead of $p.$

\bigskip

Evolutionary game theory \cite{4,5,6} does not rely on rational assumptions
but on the idea that the Darwinian process of natural selection \cite{7}
drives organisms towards the optimization of reproductive success \cite{8}.
Instead of working out the optimal strategy, the different phenotypes in a
population are associated with the basic strategies that are shaped by trial
and error by a process of natural selection or learning. The natural
selection process that determines how populations playing specific
strategies evolve is known as the replicator dynamics \cite{9,5,6,10} whose
stable fixed points are Nash equilibria \cite{2}. The central equilibrium
concept of evolutionary game theory is the notion of \textit{Evolutionary
Stable Strategy} (ESS) introduced by J. Smith and G. Price \cite{11,4}. An
ESS is described as a strategy which has the property that if all the
members of a population adopt it, no mutant strategy could invade the
population under the influence of natural selection. ESS are interpreted as
stable results of processes of natural selection.

\bigskip

Consider a large population in which a two person game $G=(S,E)$ is played
by randomly matched pairs of animals generation after generation. Let $p$ be
the strategy played by the vast majority of the population, and let $r$ be
the strategy of a mutant present in small frequency. Both $p$ and $r$ can be
pure or mixed. An \textit{evolutionary stable strategy} (ESS) $p$ of a
symmetric two-person game $G=(S,E)$ is a pure or mixed strategy for $G$
which satisfies the following two conditions%
\begin{gather}
E(p,p)>E(r,p)\text{,}  \notag \\
\text{If }E(p,p)=E(r,p)\text{ then }E(p,r)>E(r,r)\text{.}  \label{2}
\end{gather}%
Since the stability condition only concerns to alternative best replies, $p$
is always evolutionarily stable if $(p,p)$ is a strict equilibrium point. An
ESS is also a Nash equilibrium since is the best reply to itself and the
game is symmetric. The set of all the strategies that are ESS is a subset of
the NE of the game. A population which plays an ESS can withstand an
invasion by a small group of mutants playing a different strategy. It means
that if a few individuals which play a different strategy are introduced
into a population in an ESS, the evolutionarily selection process would
eventually eliminate the invaders.

\bigskip

The model used in EGT is the following: Each agent in a n-player game where
the $i^{th}$ player has as strategy space $S_{i}$ is modeled by a population
of players which have to be partitioned into groups. Individuals in the same
group would all play the same strategy. Randomly we make play the members of
the subpopulations against each other. The subpopulations that perform the
best will grow and those that do not will shrink and eventually will vanish.
The process of natural selection assures survival of the best players at the
expense of the others. The natural selection process that determines how
populations playing specific strategies evolve is known as the replicator
dynamics%
\begin{equation}
\frac{dx_{i}}{dt}=\left[ f_{i}(x)-\left\langle f(x)\right\rangle \right]
x_{i}\text{.}  \label{3}
\end{equation}%
The probability of playing certain strategy or the relative frequency of
individuals using that strategy is denoted by frequency $x_{i}$. The fitness
function $f_{i}=\sum_{j=1}^{n}a_{ij}x_{j}$ specifies how successful each
subpopulation is, $\left\langle f(x)\right\rangle
=\sum_{k,l=1}^{n}a_{kl}x_{k}x_{l}$ is the average fitness of the population,
and $a_{ij}$ are the elements of the payoff matrix $A$.

\begin{equation}
\frac{dx_{i}}{dt}=\left[ \sum_{j=1}^{n}a_{ij}x_{j}-%
\sum_{k,l=1}^{n}a_{kl}x_{k}x_{l}\right] x_{i}\text{.}  \label{4}
\end{equation}%
The replicator dynamics rewards strategies that outperform the average by
increasing their frequency, and penalizes poorly performing strategies by
decreasing their frequency. The stable fixed points of the replicator
dynamics are Nash equilibria, it means that if a population reaches a state
which is a Nash equilibrium, it will remain there.

\bigskip

Quantum games have proposed a new point of view for the solution of the
classical problems and dilemmas in game theory. Quantum games are more
efficient than classical games and provide a saturated upper bound for this
efficiency \cite{12,13,14,15,16,17}.

\section{The von Neumann Equation \& Quantum Statistical Mechanics}

An ensemble is a collection of identically prepared physical systems. When
each member of the ensemble is characterized by the same state vector $%
\left\vert \Psi (t)\right\rangle $ it is called pure ensemble. If each
member has a probability $p_{i}$ of being in the state $\left\vert \Psi
_{i}(t)\right\rangle $ we have a mixed ensemble. Each member of a mixed
ensemble\ is a pure state and its evolution is given by Schr\"{o}dinger
equation. To describe correctly a statistical mixture of states it is
necessary the introduction of the density operator%
\begin{equation}
\rho (t)=\sum_{i=1}^{n}p_{i}\left\vert \Psi _{i}(t)\right\rangle
\left\langle \Psi _{i}(t)\right\vert  \label{5}
\end{equation}%
which contains all the physically significant information we can obtain
about the ensemble in question. Any two ensembles that produce the same
density operator are physically indistinguishable. The diagonal elements $%
\rho _{nn}$ of \ the density operator $\rho (t)$ represents the average
probability of finding the system in the state $\left\vert n\right\rangle $
and its sum is equal to $1$. The non-diagonal elements $\rho _{np}$
expresses the interference effects between the states $\left\vert
n\right\rangle $ and $\left\vert p\right\rangle $ which can appear when the
state $\left\vert \Psi _{i}\right\rangle $ is a coherent linear
superposition of these states. The time evolution of the density operator is
given by the von Neumann equation%
\begin{equation}
i\hbar \frac{d\rho }{dt}=\left[ \hat{H},\rho \right]  \label{6}
\end{equation}%
which is only a generalization of the Schr\"{o}dinger equation and the
quantum analogue of Liouville's theorem.

\section{Relationships between Quantum Mechanics \& Game Theory}

In table 1 we compare some characteristic aspects of quantum mechanics and
game theory \cite{18,19,20}.

{\scriptsize Table 1}

\begin{center}
\begin{tabular}{cc}
\hline
{\scriptsize Quantum Mechanics} & {\scriptsize Game Theory} \\ \hline
{\scriptsize n system members} & {\scriptsize n players} \\ 
{\scriptsize Quantum States} & {\scriptsize Strategies} \\ 
{\scriptsize Density Operator} & {\scriptsize Relative Frequencies Vector}
\\ 
{\scriptsize Von Neumann Equation} & {\scriptsize Replicator Dynamics} \\ 
{\scriptsize Von Neumann Entropy} & {\scriptsize Shannon Entropy} \\ 
{\scriptsize System Equilibrium} & {\scriptsize Payoff} \\ 
{\scriptsize Maximum Entropy} & {\scriptsize Maximum Payoff} \\ 
&  \\ \hline
\end{tabular}
\end{center}

It is easy to realize the clear resemblances and apparent differences
between both theories and between the properties both enjoy. This was a
motivation to try to find an actual relationship between both systems.

\bigskip

It is important to note that the replicator dynamics is a vectorial
differential equation while von Neumann equation can be represented in
matrix form. If we would like to compare both systems the first we would
have to do is to try to compare their evolution equations by trying to find
a matrix representation of the replicator dynamics and this is \cite{21}%
\begin{equation}
\frac{dX}{dt}=G+G^{T}\text{,}  \label{7}
\end{equation}%
where the relative frequencies matrix $X$ has as elements $x_{ij}=\left(
x_{i}x_{j}\right) ^{1/2}$ and%
\begin{eqnarray}
\left( G+G^{T}\right) _{ij} &=&\frac{1}{2}\sum_{k=1}^{n}a_{ik}x_{k}x_{ij} 
\notag \\
&&+\frac{1}{2}\sum_{k=1}^{n}a_{jk}x_{k}x_{ji}  \notag \\
&&-\sum_{k,l=1}^{n}a_{kl}x_{k}x_{l}x_{ij}  \label{8}
\end{eqnarray}%
are the elements of the matrix $\left( G+G^{T}\right) $. From this matrix
representation we can find a Lax representation of the replicator dynamics 
\cite{21}%
\begin{equation}
\frac{dX}{dt}=\left[ \left[ Q,X\right] ,X\right]  \label{9}
\end{equation}%
and with $\Lambda =\left[ Q,X\right] $%
\begin{equation}
\frac{dX}{dt}=\left[ \Lambda ,X\right] \text{.}  \label{10}
\end{equation}%
The matrix $\Lambda $ is equal to%
\begin{equation}
(\Lambda )_{ij}=\frac{1}{2}\left[ \left( \sum_{k=1}^{n}a_{ik}x_{k}\right)
x_{ij}-x_{ji}\left( \sum_{k=1}^{n}a_{jk}x_{k}\right) \right]  \label{11}
\end{equation}%
and $Q$ is a diagonal matrix which has as elements%
\begin{equation}
q_{ii}=\frac{1}{2}\sum_{k=1}^{n}a_{ik}x_{k}\text{.}  \label{12}
\end{equation}%
It is easy to realize that the matrix commutative form of the replicator
dynamics (\ref{10}) follows the same dynamic than the von Neumann equation (%
\ref{6}) and the properties of their correspondent elements (matrixes) are
similar, being the properties corresponding to our quantum system more
general than the classical system.

\bigskip

Although a physical system is modeled and described mathematically through
quantum mechanics while a socioeconomical is modeled through game theory
both systems seem to have a similar behavior. Both are composed by $n$
members (particles, subsystems, players, states, etc.). Each member of our
systems is described by a state or a strategy which has assigned a
determined probability. The quantum mechanical system is described by a
density operator $\rho $ whose elements represent the system average
probability of being in a determined state. The socioeconomical system is
described through a relative frequencies matrix $X$ whose elements represent
the frequency of players playing a determined strategy. The evolution
equation of the relative frequencies matrix $X$ (which describes our
socioeconomical system) is given by a Lax form of the replicator dynamics
which was shown that follows the same dynamic than the evolution equation of
the density operator (the von Neumann equation).

\bigskip

Some specific resemblances between quantum statistical mechanics and
evolutionary game theory are summarized in the next table.

{\scriptsize Table 2}

\begin{center}
\begin{tabular}{cc}
\hline
{\scriptsize Quantum Statistical Mechanics} & {\scriptsize Evolutionary Game
Theory} \\ \hline
{\scriptsize n system members} & {\scriptsize n population members} \\ 
{\scriptsize Each member in the state }$\left\vert \Psi _{k}\right\rangle $
& {\scriptsize Each member plays strategy }$s_{i}$ \\ 
$\left\vert \Psi _{k}\right\rangle $ {\scriptsize with} $p_{k}\rightarrow $
\ $\rho _{ij}${\scriptsize \ } & $s_{i}${\scriptsize \ }$\ \ \rightarrow $%
{\scriptsize \ }$\ \ x_{i}$ \\ 
$\rho ,$ $\ \ \tsum_{i}\rho _{ii}{\scriptsize =1}$ & ${\scriptsize X,}$%
{\scriptsize \ \ }$\tsum_{i}x_{i}{\scriptsize =1}$ \\ 
${\scriptsize i\hbar }\frac{d\rho }{dt}{\scriptsize =}\left[ \hat{H},\rho %
\right] $ & $\frac{dX}{dt}{\scriptsize =}\left[ \Lambda ,X\right] $ \\ 
${\scriptsize S=-Tr}\left\{ {\scriptsize \rho }\ln {\scriptsize \rho }%
\right\} $ & ${\scriptsize H=-}\tsum\nolimits_{i}{\scriptsize x}_{i}\ln 
{\scriptsize x}_{i}$ \\ 
&  \\ \hline
\end{tabular}
\end{center}

In table 3 we compare the properties of the matrixes $\rho $ and $X$.

{\scriptsize Table 3}

\begin{center}
$%
\begin{tabular}{cc}
\hline
{\scriptsize Density Operator} & {\scriptsize Relative Freq. Matrix} \\ 
\hline
$\rho ${\scriptsize \ is Hermitian} & $X${\scriptsize \ is Hermitian} \\ 
${\scriptsize Tr\rho (t)=1}$ & ${\scriptsize TrX=1}$ \\ 
${\scriptsize \rho }^{2}{\scriptsize (t)\leqslant \rho (t)}$ & ${\scriptsize %
X}^{2}{\scriptsize =X}$ \\ 
${\scriptsize Tr\rho }^{2}{\scriptsize (t)\leqslant 1}$ & ${\scriptsize TrX}%
^{2}{\scriptsize (t)=1}$ \\ 
&  \\ \hline
\end{tabular}%
$
\end{center}

We can also propose the next \textquotedblleft quantization
relationships\textquotedblright 
\begin{gather}
x_{i}\rightarrow \sum_{k=1}^{n}\left\langle i\left\vert \Psi _{k}\right.
\right\rangle p_{k}\left\langle \Psi _{k}\left\vert i\right. \right\rangle
=\rho _{ii}\text{,}  \notag \\
(x_{i}x_{j})^{1/2}\rightarrow \sum_{k=1}^{n}\left\langle i\left\vert \Psi
_{k}\right. \right\rangle p_{k}\left\langle \Psi _{k}\left\vert j\right.
\right\rangle =\rho _{ij}\text{.}  \label{13}
\end{gather}%
A population will be represented by a quantum system in which each
subpopulation playing strategy $s_{i}$ will be represented by a pure
ensemble in the state $\left\vert \Psi _{k}(t)\right\rangle $ and with
probability $p_{k}$. The probability $x_{i}$ of playing strategy $s_{i}$ or
the relative frequency of the individuals using strategy $s_{i}$ in that
population will be represented as the probability $\rho _{ii}$ of finding
each pure ensemble in the state $\left\vert i\right\rangle $ \cite{21}.
Through these quantization relationships the replicator dynamics (in matrix
commutative form) takes the form of the equation of evolution of mixed
states i.e. the von Neumann equation is the quantum analogue of \ the
replicator dynamics. And also $X\longrightarrow \rho $, $\Lambda
\longrightarrow -\frac{i}{\hbar }\hat{H}$, (where $\hat{H}$ is the
Hamiltonian of the physical system) and $H(x)\longrightarrow S(\rho )$ \cite%
{22,23}.

\section{The Collective Welfare Principle \& the Quantum Understanding of
Classical Systems}

If our systems are analogous and thus exactly equivalents, our physical
equilibrium (maximum entropy) should be also exactly equivalent to our
socioeconomical equilibrium (NE or ESS). And if the natural trend of a
physical system is to a maximum entropy state, should not a socioeconomical
system trend be also to a maximum entropy state which would have to be its
state of equilibrium? Has a socioeconomical system something like a
\textquotedblleft natural trend\textquotedblright ?

\bigskip

Based on the relationships previously analyzed and specially in the
analogous behavior between quantum mechanics and game theory\textbf{,} it is
suggested the\textbf{\ }following (quantum)\textbf{\ }understanding of our
system (classical and/or socioeconomical): If in an isolated system each of
its accessible states do not have the same probability, the system is not in
equilibrium. The system will vary and will evolve in time until it reaches
the equilibrium state in where the probability of finding the system in each
of the accessible states is the same. The system will find its more probable
configuration in which the number of accessible states is maximum and
equally probable. The whole system will vary and rearrange its state and the
states of its ensembles with the purpose of maximize its entropy and reach
its equilibrium state. We could say that the purpose and maximum payoff of a
physical system is its maximum entropy state. The system and its members
will vary and rearrange themselves to reach the best possible state for each
of them which is also the best possible state for the whole system.

\bigskip

This can be seen like a microscopical cooperation between quantum objects to
improve their states with the purpose of reaching or maintaining the
equilibrium of the system. All the members of our quantum system will play a
game in which its maximum payoff is the equilibrium of the system. The
members of the system act as a whole besides individuals like they obey a
rule in where they prefer the welfare of the collective over the welfare of
the individual. This equilibrium is represented in the maximum system
entropy in where the system resources are fairly distributed over its
members. This system is stable only if it maximizes the welfare of the
collective above the welfare of the individual. If it is maximized the
welfare of the individual above the welfare of the collective the system
gets unstable and eventually it collapses (Collective Welfare Principle) 
\cite{18,19,20,21,22,23}.

\section{The Equilibrium Process called Globalization}

Lets discuss how the world process that it is called \textquotedblleft
globalization\textquotedblright\ has a behavior exactly equivalent to a
system that is tending to a maximum entropy state.

\subsection{Globalization}

Globalization represents the inexorable integration of markets,
nation-states, technologies \cite{24} and the intensification of
consciousness of the world as a whole \cite{25}.\textbf{\ }This refers to
increasing global connectivity, integration and interdependence in the
economic, social, technological, cultural, political, and ecological spheres 
\cite{26}.

\bigskip

Globalization has various aspects which affect the world in several
different ways such as \cite{26} the emergence of worldwide production
markets and broader access to a range of goods for consumers and companies
(industrial), the emergence of worldwide financial markets and better access
to external financing for corporate, national and subnational borrowers
(financial), the realization of a global common market, based on the freedom
of exchange of goods and capital (economical), the creation of a world
government which regulates the relationships among nations and guarantees
the rights arising from social and economic globalization (political) \cite%
{27}, the increase in information flows between geographically remote
locations (informational), the growth of cross-cultural contacts (cultural),
the advent of global environmental challenges that can not be solved without
international cooperation, such as climate change, cross-boundary water and
air pollution, over-fishing of the ocean, and the spread of invasive species
(ecological) and the achievement of free circulation by people of all
nations (social).

\bigskip

In economics, globalization is the convergence of prices, products, wages,
rates of interest and profits towards developed country norms \cite{28}.
Globalization of the economy depends on the role of human migration,
international trade, movement of capital, and integration of financial
markets. Economic globalization can be measured around the four main
economic flows that characterize globalization such as goods and services
(e.g. exports plus imports as a proportion of national income or per capita
of population), labor/people (e.g. net migration rates; inward or outward
migration flows, weighted by population), capital (e.g. inward or outward
direct investment as a proportion of national income or per head of
population), and technology. To what extent a nation-state or culture is
globalized in a particular year has until most recently been measured
employing simple proxies like flows of trade, migration, or foreign direct
investment, as described above. A multivariate approach to measuring
globalization is the recent index calculated by the Swiss Think tank KOF 
\cite{29}. The index measures the three main dimensions of globalization:
economic, social, and political.

\bigskip

Maybe the firsts of these so called states-nations, communities,
\textquotedblleft unions\textquotedblright , common markets, etc. were the
Unites States of America and the USSR (now the Russian Federation). Both
consists in a set or group of different nations or states under the same
basic laws or principles (constitution), policies, objectives and an economy
characterized by a same currency. Although each state or nation is a part of
a big community each of them can take its own decisions and choose its own
way of government, policies, laws and punishments (e.g. death penalty) but
subject to a constitution (which is no more than a \textquotedblleft common
agreement\textquotedblright )\ and also subject to the decisions of the
\textquotedblleft congress\textquotedblright\ of the community which
regulates the whole and the individual decisions of the parts. The United
States of America consists in 50 states and a federal district. It also has
many dependent territories located in the Antilles and Oceania. The currency
in The United States is the\ Dollar. The Russian Federation consists in a
big number of political subdivisions (88 components). There are 21 republics
inside the federation with a big degree of autonomy over most of the
aspects. The rest of territory consists in 48 provinces known as \'{o}blast
and six regions (kray), between which there are 10 autonomic districts and
an autonomic \'{o}blast and 2 federal cities (Moscow and San Petersburg).
Recently, seven federal districts have been added. The currency in Russia is
the\ Rublo \cite{26}.

\bigskip

The European Union stands as an example that the world should emulate by its
sharing rights, responsibilities, and values,\ including the obligation to
help the less fortunate. The most fundamental of these values is democracy,
understood to entail not merely periodic elections, but also active and
meaningful participation in decision making, which requires an engaged civil
society, strong freedom of information norms, and a vibrant and diversified
media that are not controlled by the state or a few oligarchs. The second
value is social justice. An economic and political system is to\ be judged
by the extent to which individuals are able to flourish and realize their
potential. As individuals, they are part of an ever-widening circle of
communities, and they can realize their potential only if they live in
harmony with each other. This, in turn, requires a sense of responsibility
and solidarity \cite{30}.

\bigskip

The meeting of 16 national leaders at the second East Asia Summit (EAS) on
the Philippine island of Cebu in January 2007 offered the promise of the
politically fractious but economically powerful Asian mega-region one day
coalescing into a single meaningful unit \cite{31}.

\bigskip

Seth Kaplan has offered the innovative idea of a West African Union to help
solve West Africa's deep-rooted problems. The 15 West African countries
stretching from Senegal to Nigeria face some of the worst problems of the
developing world: \ Pervasive inter-group conflict; borders that fail to
reflect the cultural \ landscape; weak national cohesion; corrupt officials
and impotent institutions; a dearth of skilled workers exacerbated by brain
drain; poor\ investment climates; and AIDS. Britain's Department for
International Development considers ten of the fifteen countries fragile.
Seventy-five percent of the area's people live under governments that cannot
deliver many of the most basic services---including, in many cases,
security. More than 25,000 peacekeepers are needed to maintain a fragile
peace in the region's war zones. Conflicts spill easily across borders, as
do refugees, arms, and instability \cite{32}.

\bigskip

In South America has been proposed the creation of a Latin-American
Community which is an offer for the integration, the struggle against the
poverty and the social exclusion of the countries of Latin-America. It is
based on the creation of mechanisms to create cooperative advantages between
countries that let balance the asymmetries between the countries of the
hemisphere and the cooperation of funds to correct the inequalities of the
weak countries against the powerful nations. The economy ministers of
Paraguay, Brazil, Argentina, Ecuador, Venezuela and Bolivia agreed in the
\textquotedblleft Declaraci\'{o}n de Asunci\'{o}n\textquotedblright\ to
create the Bank of the South and invite the rest of countries to add to this
project. The Brazilian economy minister Mantega stand out that the new bank
is going to consolidate the economic, social and politic block that is
appearing in South America and now they have to point to the creation of a
common currency. Recently, Uruguay has also accepted the offer of the
creation of the bank and the common currency and is expected that more
countries add to this offer \cite{33}.

\subsection{The Equilibrium Process}

After analyzing our systems we conclude that a socioeconomical system has a
behavior exactly equivalent that a physical system. Both systems evolve in
analogous ways and to analogous points. A system where its members are in
Nash Equilibrium (or ESS) is exactly equivalent to a system in a maximum
entropy state. The stability of the system is based on the maximization of
the welfare of the collective above the welfare of the individual. The
natural trend of a physical system is to a maximum entropy state, should not
a socioeconomical system trend be also to a maximum entropy state which
would have to be its state of equilibrium? Has a socioeconomical system
something like a \textquotedblleft natural trend\textquotedblright ?

\bigskip

From our analysis a population can be represented by a quantum system in
which each subpopulation playing strategy $s_{i}$ will be represented by a
pure ensemble in the state $\left\vert \Psi _{k}(t)\right\rangle $ and with
probability $p_{k}$. The probability $x_{i}$ of playing strategy $s_{i}$ or
the relative frequency of the individuals using strategy $s_{i}$ in that
population will be represented as the probability $\rho _{ii}$ of finding
each pure ensemble in the state $\left\vert i\right\rangle $ \cite{21}.
Through these quantization relationships the replicator dynamics (in matrix
commutative form) takes the form of the equation of evolution of mixed
states i.e. the von Neumann equation is the quantum analogue of \ the
replicator dynamics. And also $X\longrightarrow \rho $, $\Lambda
\longrightarrow -\frac{i}{\hbar }\hat{H}$, (where $\hat{H}$ is the
Hamiltonian of the physical system) and $H(x)\longrightarrow S(\rho )$ \cite%
{22,23}.

\bigskip

Our now \textquotedblleft quantum statistical\textquotedblright\ system
composed by quantum objects represented by quantum states which represent
the strategies with which \textquotedblleft players\textquotedblright\
interact would be characterized by certain interesting physical parameters
like temperature, entropy and energy. Moreover, we could define an entropy,
an energy and a temperature in every socioeconomical system but better and
actually our socioeconomical system would be characterized by an entropy, an
energy, a temperature and would have a similar or analogous behavior to a
physical system. From entropy we can analyze our system from two different
points of view through quantum information theory and statistical mechanics 
\cite{23,18,19,20}.

\bigskip

In this statistical mixture of ensembles (each ensemble is characterized by
a state and each state has assigned a determined probability) its natural
trend is to its maximum entropy state. If each of its accessible states do
not have the same probability, the system will vary and will evolve in time
until it reaches the equilibrium state in where the probability of finding
the system in each of the accessible states is the same and its number is
maximum. In this equilibrium state or maximum entropy state the system
\textquotedblleft resources\textquotedblright\ are fairly distributed over
its members. Each ensemble will be equally probable and will be
characterized by a same temperature and will be in a stable state.

\bigskip

Socioeconomically and based on our analysis, our world could be understood
as a statistical mixture of \textquotedblleft ensembles\textquotedblright\
(countries for example). Each of these ensembles are characterized by a
determined state and a determined probability. But more important, each
\textquotedblleft country\textquotedblright\ is characterized by a specific
\textquotedblleft temperature\textquotedblright\ which is a measure\textbf{\ 
}of the socioeconomical activity of that ensemble. That temperature is
related with the activity or with the interactions between the members of
the ensemble. The system will evolve naturally to a maximum entropy state.
Each pure ensemble of this statistical mixture will vary and accommodate its
state until get the \textquotedblleft thermal equilibrium\textquotedblright\
first with its nearest neighbors creating new big ensembles characterized
each of them by a same temperature. Then with the time, these new big
ensembles will seek its \textquotedblleft thermal
equilibrium\textquotedblright\ with its nearest neighbors\ creating new
bigger ensembles but less in number. The system will continue evolving
naturally in time until the whole get a state characterized by a same
\textquotedblleft temperature\textquotedblright .

\bigskip

This behavior is very similar to what have been called globalization and
actually the process of equilibrium that is absolutely equivalent to a
system that is tending to a maximum entropy state is the actual
globalization. This analysis predicts the apparition of big common
\textquotedblleft markets\textquotedblright\ or (economical, political,
social, etc.) communities of countries (European Union, Asian Union,
Latin-American Community, African Union, Mideast Community, Russia and USA)
and strong common currencies (dollar, euro, yen, inti, etc.). The little and
poor economies eventually will finish unavoidably absorbed by these
\textquotedblleft markets\textquotedblright\ and these currencies. If this
process continues these markets or communities will find its
\textquotedblleft equilibrium\textquotedblright\ by decreasing its number
until there will be a \ moment in where there exists a big common community
or market and a same common currency around the world.

\section{Conclusions}

Although both systems analyzed are described through two theories apparently
different (quantum mechanics and game theory) both are analogous and thus
exactly equivalents. A socioeconomical system has a behavior exactly
equivalent that a physical system. Both systems evolve in analogous ways and
to analogous points. The quantum analogue of the replicator dynamics is the
von Neumann equation. A system where its members are in Nash Equilibrium (or
ESS) is exactly equivalent to a system in a maximum entropy state. The
stability of the system is based on the maximization of the welfare of the
collective above the welfare of the individual. The natural trend of both
systems is to its maximum entropy state which is its state of equilibrium.
The results of this work show that the \textquotedblleft
globalization\textquotedblright\ process has a behavior exactly equivalent
to a system that is tending to a maximum entropy state and predicts the
apparition of big common markets and strong common currencies that will find
its \textquotedblleft equilibrium\textquotedblright\ by decreasing its
number until they get a state characterized by a big common community or
market and a same common currency around the world.

\end{document}